# An Extended Radio Counterpart of TeV J2032+4130?

Y. M. Butt[1], J. A. Combi[2], J. Drake[1], J. P. Finley[3], A. Konopelko[3], M. Lister[3], J. Rodriguez[4], D. Shepherd[5]

[1]*Harvard-Smithsonian Center for Astrophysics, 60 Garden St., Cambridge, MA 02138, USA*
[2]*Departamento de Física (EPS), Universidad de Jaén, Campus Las Lagunillas s/n, 23071 Jaén, SPAIN*
[3]*Department of Physics, Purdue University, West Lafayette, IN 47907, USA*
[4]*CEA Saclay, DSM/DAPNIA/SAp,F-91191 Gif sur Yvette, FRANCE*
[5]*NRAO, P.O. Box O, Socorro, NM 87801-0387, USA*

**Abstract.** We carried out a 5-pointing mosaic observation of TeV J2032+4130 at 1.4 and 4.8 GHz with the VLA in April of 2003. The analysis of the 4.8GHz data indicate weak wispy shell-like radio structure(s) which are at least partially non-thermal. The radio data are compatible with one or more young supernova remnants or perhaps the signature of large-scale cluster shocks in this region induced by the violent action of the many massive stars in Cyg OB2.

**Keywords:** Enter Keywords here.
**PACS:** Replace this text with PACS numbers;

## INTRODUCTION

TeV J2032+4130 is a steady and extended very high energy (VHE) gamma-ray emitter located within the angular extent of the Cygnus OB2 association. When it was reported by the HEGRA collaboration in 2002 [1] it was remarkable in that it was the first unidentified VHE source – and it still remains unidentified. We have previously suggested that this TeV source may be related to an outlying sub-group of powerful OB stars belonging to Cyg OB2 (Fig 1, from [2]). In fact, the clustering of point-like X-ray sources we saw in a moderately deep (~50ksec) Chandra exposure lends some credence to this scenario (Fig. 2, from [3]), but adjacent Chandra fields are needed to confirm this. The recent detection of the Westerlund 2 stellar cluster by the HESS collaboration supports such a thesis[1], as does recent theoretical work [4].

## VLA D-configuration Radio Observations

On April 29 2003 we carried out a mosaic observation of the TeV J2032+4130 source region with the VLA[2]. The array was in the D-configuration and the flux calibrator was 3C48 and the gain calibrator, J2007+404. The source fields were observed with 3.5 to 6' spacing in a 5-point pattern, the primary beam of the VLA at 4.85 GHz being about 9'.

The resulting 6cm data were reduced with the Common Astronomy Software Applications (CASA) software and imaged with Astronomical Information Processing Software (AIPS++). The final mosaic image was weighted with robust weighting that provided an optimal compromise between point source sensitivity and minimum noise levels. The image was gridded to form a single mosaic and then deconvolved with a CLEAN algorithm that allowed multiple scales to be used for the clean components to maximize sensitivity to extended structure while still

---

[1] HESS J1023–575 is coincident with Westerlund 2 see:
http://www.mpi-hd.mpg.de/hfm/HESS/public/som/Som_12_06.htm

[2] NRAO VLA proposal AB1075 (PI: Butt) Observed April 29, 2003. 10 hours total in a five-pointing exposure at 1.4 and 4.8GHz. The NRAO is a facility of the NSF operated under cooperative agreement by Associated Universities, Inc

preserving the image flux density. The resulting image has uniform sensitivity across the mosaic field, however the flux density decreases toward the edge of the field as the primary beam response decreases.

The flux was measured in an image that was corrected for the mosaic primary beam response so the flux density was constant while the noise increased toward the edge of the mosaic field. The diffuse structure has a flux density of approximately 225 mJy at 6cm. In comparing the 6cm data with the 20cm data [5], it is evident that the western region of this extended shell-like structure is predominantly non-thermal.

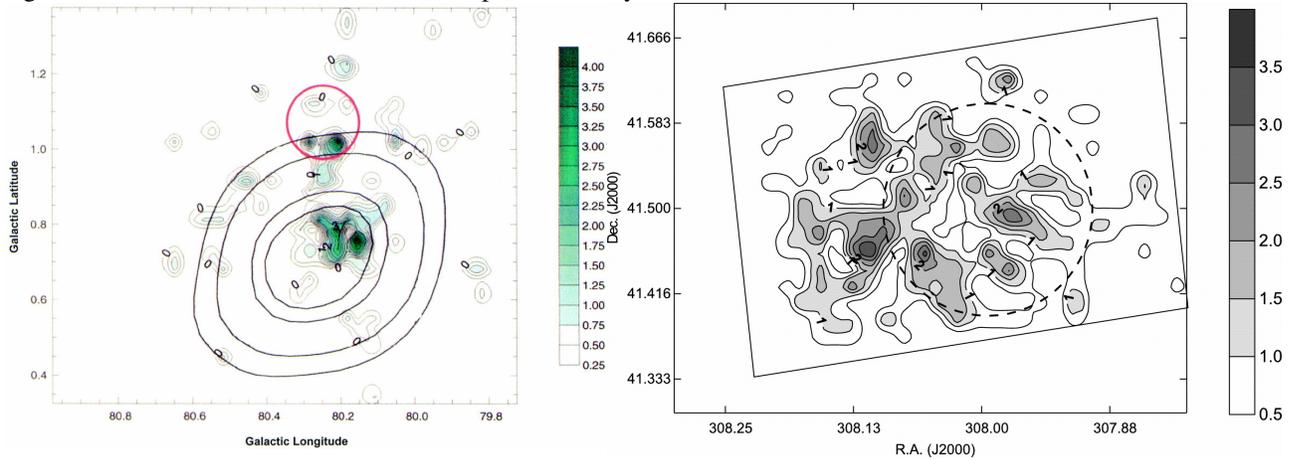

**Figure 1. (a) The OB-stellar surface density – the TeV source region is shown in red.** *From Ref. 1* **(b) The surface density of the 220 point-like X-ray sources detected by Chandra in the field. The circle marks the TeV source region.** *From Ref. 2*

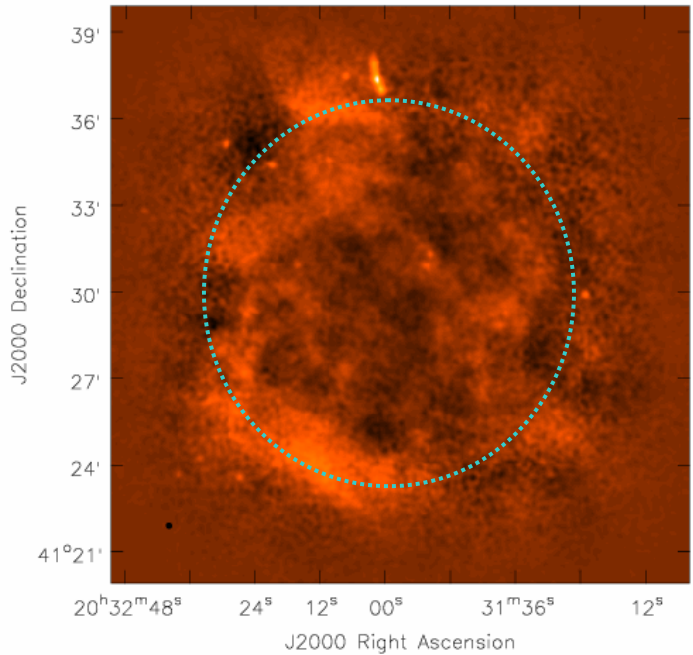

**Figure 2: A 4.8GHz VLA mosaic image of the five pointings towards TeV J2032+4130. The total diffuse flux density is ~225 mJy. The flux density in the image ranges from -0.3 mJy/beam to 0.69 mJy/beam**

# REFERENCES

1. Aharonian, F., et al. 2002, A&A, 393, L37
2. Butt, Y., et al., 2003, ApJ, 597, 494
3. Butt, Y., et al., 2006, ApJ, 643, 238
4. Domingo-Santamaria, E., & Torres, D., 2006, A&A 448, 613
5. Paredes, J. M., et al., ApJ 2007, 654, L135